\documentclass[sigconf,screen]{acmart}

\usepackage{url}
\usepackage{booktabs}
\usepackage{algorithmic}
\usepackage{graphicx}
\usepackage{textcomp}
\usepackage{xcolor}
\usepackage{hyperref}
\def\BibTeX{{\rm B\kern-.05em{\sc i\kern-.025em b}\kern-.08em
    T\kern-.1667em\lower.7ex\hbox{E}\kern-.125emX}}
\usepackage{xspace}
\usepackage{cleveref}
\usepackage[caption=false]{subfig}  
\usepackage{microtype}

\newcommand{\ie}{\emph{i.e.,}\xspace}
\newcommand{\eg}{\emph{e.g.,}\xspace}

\newcommand{\toolname}{\emph{CodeChangeEmbedder}\xspace}
\newcommand{\yinmodel}{\emph{BaseModel}\xspace}

\begin{document}

\title[Unsupervised Learning of General-Purpose Embeddings for Code Changes]{Unsupervised Learning of General-Purpose \\ Embeddings for Code Changes}

\author{Mikhail Pravilov}
\affiliation{
    \institution{Higher School of Economics}
    \city{Saint Petersburg}
    \country{Russia}
}
\email{mepravilov@edu.hse.ru}

\author{Egor Bogomolov}
\affiliation{
    \institution{JetBrains Research}
    \institution{Higher School of Economics}
    \city{Saint Petersburg}
    \country{Russia}
}
\email{egor.bogomolov@jetbrains.com}

\author{Yaroslav Golubev}
\affiliation{
    \institution{JetBrains Research}
    \city{Saint Petersburg}
    \country{Russia}
}
\email{yaroslav.golubev@jetbrains.com}

\author{Timofey Bryksin}
\affiliation{
    \institution{JetBrains Research}
    \institution{Higher School of Economics}
    \city{Saint Petersburg}
    \country{Russia}
}
\email{timofey.bryksin@jetbrains.com}

\begin{abstract}

Applying machine learning to tasks that operate with code changes requires their numerical representation. In this work, we propose an approach for obtaining such representations during pre-training and evaluate them on two different downstream tasks --- applying changes to code and commit message generation. During pre-training, the model learns to apply the given code change in a correct way. This task requires only code changes themselves, which makes it unsupervised. In the task of applying code changes, our model outperforms baseline models by 5.9 percentage points in accuracy. As for the commit message generation, our model demonstrated the same results as supervised models trained for this specific task, which indicates that it can encode code changes well and can be improved in the future by pre-training on a larger dataset of easily gathered code changes.

\end{abstract}



\maketitle

\section{Introduction}\label{sec:introduction}

In many software engineering tasks, researchers analyze not only the source code itself, but also the way code changes. For example, in a well-known task of generating a message that describes a commit in a version control system, the main focus is on the change itself~\cite{jiang2017automatically, buse2010automatically}. Other tasks that require code transformations are bug fixing~\cite{long2016automatic, tufano2018empirical}, code refactoring~\cite{fokaefs2011jdeodorant, tufano2019learning},
and others. 
There are also problems that require classification of code changes, \eg stable patch prediction~\cite{hoang2020cc2vec} or security-relevant patch prediction~\cite{lozoya2019commit2vec}. 

In order to apply machine learning methods to any objects, including code changes, one needs to represent them as numeric vectors.
Recent studies in this area developed several approaches based on deep neural networks. They allow building distributed vector representations (so called embeddings) that automatically capture the most relevant features of the encoded objects. Following this line of work, embeddings of code changes can be built during the process of training a neural network to solve a specific software engineering task (\eg stable patch prediction~\cite{hoang2020cc2vec}). In this case, the embeddings are task-specific by their nature. Other researchers suggest methods that are able to build general-purpose embeddings that can be later fine-tuned for a particular downstream task~\cite{hoang2020cc2vec, yin2018learning}. 

Existing works can be classified not only by the type of embeddings they build, but also by how their models are trained: supervised~\cite{lozoya2019commit2vec, hoang2020cc2vec} or unsupervised~\cite{yin2018learning}. Supervised in this case means any technique that needs labeled data of any kind. For example, if a model is designed for the commit message generation task and requires commit messages in the dataset to train, this makes an approach supervised. 
Collecting a large labeled dataset often requires a lot of human effort.  
On the contrary, an unsupervised approach that solves the same task would require a collection of only code changes themselves, which are much easier to get.

In this paper, we propose a new unsupervised approach to building general-purpose embeddings of code changes and solving software engineering tasks that employ them. The approach is based on an unsupervised pre-training of a deep neural network to build distributed edit representations.\footnote{In this work, we use the terms \textit{code edits}, \textit{code modifications}, and \textit{code changes} interchangeably.} To apply them to a particular task that relates to code changes, the user only needs to add some additional neural layers to the proposed model to fine-tune these edit representations for the specific task. 

We also aim to research whether general-purpose embeddings built in an unsupervised manner can be effectively used in practical software engineering tasks. We evaluate the approach on two separate tasks: applying changes to code and commit message generation. In the task of applying changes to code, we show how to make edit representations more generalized and report better performance compared to other neural models proposed in prior work for this task. In the commit message generation task, our model achieves the same results as neural machine translation models trained specifically for this task~\cite{jiang2017automatically, liu2019generating}. This indicates that the constructed embeddings of code changes contain enough information to describe them in a natural language.

The main contributions of this paper are:
\begin{itemize}
    \item We propose a novel approach for solving software engineering tasks related to code changes. It consists of unsupervised pre-training of a neural network to build general-purpose embeddings of code changes and further use of these embeddings to solve various software-engineering tasks. 
    \item Based on the proposed approach, we implement a model called \toolname. We make all the data and code publicly available, including the replication packages not only for our work but also for other papers that we use as baselines~\cite{yin2018learning, jiang2017automatically, liu2019generating}, which makes reproducibility and direct comparison with them easier for other researchers in this area. You can find all the data in the replication package: \url{https://zenodo.org/record/5082684}.
    \item We conduct an evaluation on two software engineering tasks to show the applicability of the proposed unsupervised approach. The first task is applying changes to code, where our model improves the accuracy by $5.9$ percentage points compared to another unsupervised approach~\cite{yin2018learning} and is able to build more generalized representations of code changes. The second task is commit message generation, where \toolname has the same performance as state-of-the-art supervised models.
\end{itemize}

\vspace{-0.2cm}
\section{Background}\label{sec:background}

Recent works in the natural language processing (NLP) field employ pre-training models on large corpora of texts in an unsupervised manner. Radford et al.~\cite{radford2018improving} offer to pre-train a deep neural network with the Transformer architecture~\cite{vaswani2017attention} on the task of predicting the next word given the preceding context. Devlin et al.~\cite{devlin2018bert} propose an approach called BERT which consists of pre-training a Transformer as well, but on a different task: predict a masked word given words before and after it.
Driven by these advances in NLP, a lot of works have been published recently that aim to vectorize a code snippet in similar ways, including unsupervised approaches. A comprehensive list has been recently assembled by Chen et al.~\cite{chen2019literature}. For example, Kanade et al.~\cite{kanade2019pre} pre-train BERT on source code and evaluate the obtained contextual embeddings on five classification tasks. 

A separate and necessary task is building representations not of the source code itself but of its changes. In this field, most of the works focus on building task-specific vector representations of changes. 
For example, Jiang et al.~\cite{jiang2017automatically} gather a dataset of Git commits and their messages and train a neural machine translation model with the attention mechanism to generate commit messages from code changes. Liu et al.~\cite{liu2019generating} add the copying mechanism to the model of Jiang et al. to improve its performance.

There are also works that build explicit general-purpose embeddings of code changes that can be used in various software engineering tasks. Hoang et al.~\cite{hoang2020cc2vec} propose a neural architecture where they leverage structural information from the code change, while still treating it as a sequence of tokens. 
The authors aggregate the information from individual tokens into a single vector using the attention mechanism and LSTM cells~\cite{hochreiter1997long}. 
During the training phase, the model has to predict for each token in the vocabulary whether it is contained in a commit message or not. We consider this approach as supervised because it requires each change to have a short description of it in a natural language for training, which in this case acts as a label.

To the best of our knowledge, there exists only one work that suggests an unsupervised approach to learn embeddings of code changes. Yin et al.~\cite{yin2018learning} offer a new unsupervised training objective that requires only the source code before and after the change: based on these two fragments of code, the authors construct an edit sequence, and the model trains to apply the constructed edit to the code before the change to generate the code after the change. The suggested neural network employs a classical encoder-decoder architecture. The authors conducted experiments to see if the obtained embeddings could be grouped in semantically meaningful clusters and how well the representations are generalized and transferred from one context to another. 

\vspace{-0.2cm}
\section{Proposed Approach}\label{sec:model}

In this paper, we propose a novel approach to solving practical tasks related to code changes. We suggest to pre-train a deep neural model aimed to build general-purpose embeddings of code changes and then further fine-tune it for particular software engineering tasks. The pre-training task is to apply edits to source code. This task fits perfectly since it requires only a collection of code changes themselves. This makes the pre-training step completely unsupervised and allows to use any dataset of code changes, while platforms like GitHub allow collection of such data at large scale.

\vspace{-0.2cm}

\subsection{Model}

The model we propose to use is based on an existing model by Yin et al.~\cite{yin2018learning}. In this paper, we will also refer to their model as \yinmodel. The goal of both approaches is to learn distributed vector representations of code changes. As input, the models receive two fragments of code, before and after the change. These two fragments are treated as sequences of tokens and they are used to construct an edit sequence.

\begin{figure}[htbp]
\centerline{\includegraphics[width=\columnwidth, keepaspectratio,]{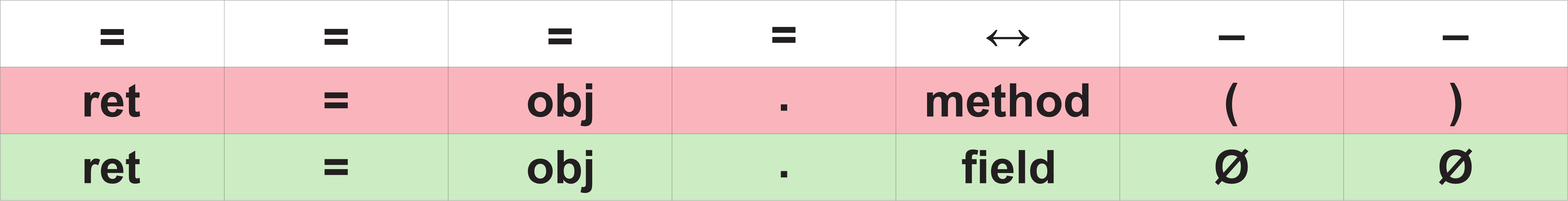}}
\caption{An example of edit sequence used by Yin et al.~\cite{yin2018learning}.}
\label{fig:edit_sequence}
\end{figure}

An edit sequence is a sequence of columns, which is constructed by applying a deterministic diffing algorithm based on the Levenshtein distance~\cite{levenshtein1966binary} to the two input sequences of tokens. In each column, the first row indicates the action that happened with the token: replaced ($\leftrightarrow$), added ($+$), deleted ($-$), or unchanged ($=$). The second row contains a token before the change and the third row contains a token after the change. In the case of additions or deletions, padding symbols are used ($\varnothing$). All special symbols are treated as ordinary tokens later. An example of such alignment is shown in \Cref{fig:edit_sequence}. 

The model trains to generate the modified code snippet given the original code fragment and the change sequence. The training is guided to minimize the difference between the generated code and the ground truth (real code after the edit from the dataset). Such tasks are usually solved by employing encoder-decoder architectures common in neural machine translation (NMT). Both encoders and decoders are neural networks, with encoders being networks that produce vector representations of the input, and decoders being networks that transform representations into output sequences.

The overall architecture of the proposed model is depicted in \Cref{fig:model_architecture}. In our model, we have two encoders: one to encode the code before the edit, and the second one to encode the edit itself. Thus, after the encoding step, we have two separate vectors: one for the code and one for the edit. Given them, the decoder has to produce a sequence of tokens that represents the code after the change. Other components that are used in our model are attention and copying~\cite{vinyals2015pointer} mechanisms. The aim of the attention mechanism is to highlight the most relevant tokens in the input sequence to make the decoder pay more attention to them during decoding steps. To estimate how relevant the input tokens are, the model calculates non-negative weights that sum up to 1 in total. The more important the token is, the closer its weight is to 1. The copying mechanism is necessary to deal with out-of-vocabulary tokens --- the tokens that the model does not recognize and is unable to generate. Similar to the attention mechanism, for each input token, a weight is calculated, only now it is treated as a probability to copy this token into the output sequence. This helps to generate tokens that are not in the model's vocabulary, but that are present in the input code fragment.

In our model, we use LSTM with attention and copying mechanisms as a decoder. The LSTM state is initialized not only with the representation of code before the change but also with the edit representation. Also, the edit representation vector is fed into LSTM at each decoding step. 

\subsection{Proposed Modifications}

We propose to change the way how edits are represented in the model compared to \yinmodel. One important observation is that the decoder receives not only the edit representation but also the representation of the code before the change. This means that the information about the unchanged tokens in edit representations becomes redundant. Moreover, unchanged tokens can prevent representations of structurally similar edits to group together because of the different contexts they have. This can lead to the problem when similar edits cannot be applied to code fragments with different contexts. 

Based on this observation, in our edit sequences we leave only the changed tokens and remove those that remain unchanged during code modification. This should help to make the edit representation vectors more generalized and vectors of structurally similar changes closer to each other.

\begin{figure}[t]
\centerline{\includegraphics[width=\columnwidth, keepaspectratio,]{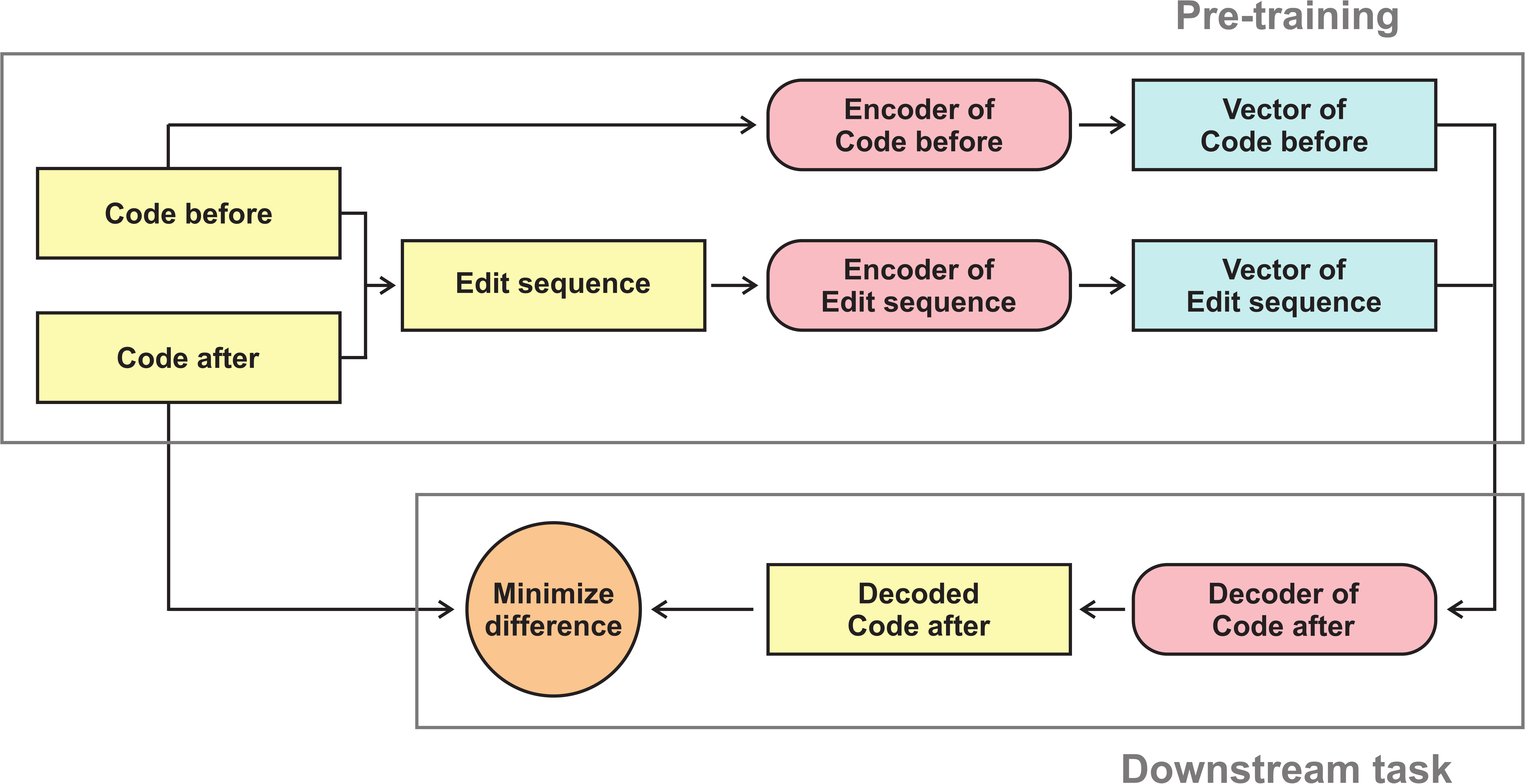}}
\caption{The architecture of the model. For fine-tuning the model for a particular task, the decoder should be replaced with new neural layers that output the predictions.}
\label{fig:model_architecture}
\end{figure}

Also, Yin et al. consider their approach mostly as a model that is capable of applying edits to text or source code. In our work, we propose to take a different look at this. By changing the way edit sequences are built, we are able to retrieve more generalized representations of edits. Thus, the embeddings of edits can then be used in any downstream task that requires working with code changes. This includes not only direct applying of edits such as bug-fixing or code refactoring, but also commit message generation, classification of changes, and others.

The main advantage of the proposed approach is that the training objective is completely unsupervised. We offer to pre-train the proposed model to build the edit representations on a large corpus of unlabeled data and then fine-tune it for a specific task on a small corpus of high-quality labeled data. 
Parts of the model that should be left for fine-tuning are shown at the top of the \Cref{fig:model_architecture} and are labelled as \textit{Pre-training}. When the pre-training is finished, we can add new neural layers instead of the old decoder. These layers can be of any kind, including fully connected layers for classification tasks or recurrent neural layers for generation tasks. The weights of the new neural layers will be fine-tuned for a specific task.
Taking into consideration the success of unsupervised pre-training in the NLP domain~\cite{devlin2018bert, radford2019language}, we see a great potential of \toolname in tasks that relate to code changes. 

\section{Evaluation}\label{sec:evaluation}

We evaluate our approach on two popular software engineering tasks. The first one is applying changes to code. Specifically, a model should generate a changed code fragment given the initial fragment and the edit representation as input. This task is very similar to the one that we use at the pre-training step, but here the model does not know the code after the change, therefore, in this evaluation we cannot build edit sequences as before. Instead, we use manual labeling. Each code snippet has a label that identifies how this code snippet should be changed structurally. According to this label, we pick an edit sequence from the training set which has the same label and represents a similar change. In this experimental setup, the primary goal of the evaluation is to compare \toolname and \yinmodel in terms of building distributed representations of code changes that are consistent with the manual labeling.

Secondly, we evaluate the proposed approach on the commit message generation task. In essence, this is the task of describing a change in a natural language. Given a commit, a model must generate a text that describes it. Several existing works claimed that a lot of developers neglect writing commit messages due to the lack of time~\cite{buse2010automatically, jiang2017automatically}, however, these messages are very important for the maintenance of projects and their future development~\cite{buse2010automatically}. Automatic generation of decent commit messages could improve the quality of software and simplify the work of developers.

\subsection{Applying Changes to Code}\label{sec:application-of-code-changes}

For our experiments, we took two datasets by Tufano et al.~\cite{tufano2018empirical, tufano2019learning}:

\textbf{BugFixes}. The dataset consists of bug fixing examples written in Java and mined from every public GitHub event stored in the GitHub Archive~\cite{githubarchive} between March 2011 and October 2017. Such a sampling includes a huge amount of commits in different projects hosted on GitHub and allows to cover diverse project topics. The authors filtered this dataset to contain only methods no longer than 50 tokens, which left them with 58,350 unlabeled samples.

\textbf{Labeled}. This dataset contains 388 samples selected by Tufano et al. from code changes mined from three Gerrit\footnote{Gerrit, a platform for code review: \url{https://www.gerritcodereview.com/}} projects written in Java. The topics of these projects are rather diverse: Android, virtualization, and code collaboration. 
The authors manually labeled the changes with the aim to cluster the changes semantically and analyze them. As a result, the authors obtained 64 classes of code changes. Because of the manual effort, \textit{Labeled} is much smaller than the previous  dataset, but is perfectly suitable for testing.

Both datasets consist of source code of Java methods before and after the change. To reduce the vocabulary size, Tufano et al. applied a canonicalization technique. For example, they replaced rarely used identifiers with generic names (\eg ``VARIABLE\_1", ``TYPE\_2", ``METHOD\_3"). The canonicalization process takes into account both versions of the method, so that identical identifiers get identical generic names. The code of the methods represented as sequences of tokens does not exceed 50 tokens.

We compare \toolname with the model from the work of Tufano et al.~\cite{tufano2018empirical} and \yinmodel~--- the model by Yin et al. that was described in detail in \Cref{sec:model}. The model of Tufano et al. is a classic neural machine translation model with the attention mechanism. The authors used LSTMs for both encoder and decoder. 

We trained all the models on the \textit{BugFixes} dataset, which was split into training, validation, and test sets in the 8:1:1 ratio, respectively. We used the exact same splitting that was performed by Tufano et al. in their work. 
All the models have decoders as the last step of their pipelines. To provide fair comparison, we used beam search~\cite{tillmann2003word} with the width of 50 for decoding, the same value that Tufano et al. used in their work~\cite{tufano2018empirical}.

After training, we evaluated the models on the \textit{Labeled} dataset. For \yinmodel and \toolname, we randomly chose a representative example in each of the change classes (random selection was performed to avoid the bias of sampling). We then applied the edit representation of the representative example to other code examples in the same class. The idea behind this experiment is that if we have a group of similar changes, they all should have similar edit representation vectors, thus the same edits should be successfully applied to other code fragments within this group. The experiment shows how well edit representations correspond to the human judgment, which we consider as ground truth in this evaluation.

We used the model of Tufano et al. as a baseline, since it does not use any information except the code before the change. During the training, it simply memorizes that a particular code snippet must be transformed into another particular code snippet. On the other hand, \toolname and \yinmodel transform the code according to the encoded edit and as a result have more information during the prediction step than the baseline. Therefore, we hypothesize that they should be more accurate in their predictions.

\vspace{-0.3cm}

\subsection{Commit Message Generation}\label{sec:commit-message-generation}

For the commit message generation task, we conduct our experiments on the data collected by Jiang et al.~\cite{jiang2017automatically}, since it is the most popular dataset used in prior works for this task. The dataset includes commits from the 1,000 most starred Java projects on GitHub. 1,000 projects is a large enough sample that includes projects from different domains, which allows us to avoid bias caused by the prevalence of specific topics. If a commit message has more than one sentence in it, only the first sentence is left because usually it summarizes the whole commit message~\cite{gu2016deep}. The code of a commit is represented as a sequence of tokens obtained from the output of \texttt{git diff}~\cite{gitdiff}. It shows the difference between versions of files on the level of code lines, \ie indicates which lines were removed, added, changed, or left unchanged. The lengths of all code sequences in this dataset are between 50 and 100 tokens, and the lengths of all commit messages are less than 30 tokens. This dataset contains 32,205 samples in total, in our paper we refer to it as \textit{Original}.

Since our unsupervised pre-training works only with changes, we also introduced a \textit{Filtered} dataset. First of all, it removes commits where entire files were added or deleted, in which case we do not have the code fragment before or after the change accordingly. Also, to lower the scope of the change, it removes commits where several files were changed. This leaves us with approximately 80\% of the data, resulting in the dataset with 25,555 samples that was still sufficient enough to train a neural network. This data was split into two equal parts: the first one was used to pre-train the network and the second one was used for fine-tuning on the commit message generation task. We decided not to mine additional data to preserve the homogeneity of the dataset and provide a fair comparison with other models.

We compare our model with the approach by Jiang et al.~\cite{jiang2017automatically} and with the approach by Liu et al.~\cite{liu2019generating}, since these works propose supervised approaches to commit message generation, while ours is unsupervised. Jiang et al. suggested to use a neural machine translation model with the attention mechanism to translate changes in Java code into natural language. Liu et al. improved the model by adding the copying mechanism,
which improved the performance compared to the model of Jiang et al. 

Firstly, we conduct an evaluation on the \textit{Original} dataset to compare with the results from the respective papers~\cite{jiang2017automatically, liu2019generating}. During this evaluation our model was fine-tuned on the same data it was pre-trained on. The models of Jiang et al. and Liu et al. were trained as usual on the training data once. 

Secondly, we conduct an evaluation on the \textit{Filtered} dataset to see how the suggested approach behaves when the pre-training and fine-tuning data actually differ. We split the \textit{Filtered} dataset in two halves: the first part was used only for pre-training of \toolname, the second part of the data was available for all the compared models. \toolname used this data for fine-tuning, while other models were trained on this part as usual. For pre-training, the model was pre-trained with an unsupervised training objective described in \Cref{sec:model} without seeing the commit messages. After that, the weights of the encoders were frozen and the old decoder was replaced with a new one. The new decoder was fine-tuned on the second half of the dataset to extract the information needed to generate commit messages from the general-purpose vector representations of changes.

To ensure that the results are not biased due to a specific train/test split, we conducted a 10-fold cross-validation. In the case of the \textit{Filtered} dataset, its first part is split into training, validation, and test sets only once for all 10 evaluations. All splits are made with a 8:1:1 ratio for training, validation, and test sets. We used BLEU~\cite{papineni2002bleu} as a metric in these experiments since we compare two texts in a natural language.

\section{Results}\label{sec:results-discussion}

\subsection{Applying Changes to Code}

The evaluation results of applying changes to code are presented in \Cref{tab:evaluation_of_model_modifications}. It can be noted that the accuracy values for all the models are very low, however, this is because the task itself is very difficult. The models have to generate a lot of consecutive tokens and an error in any of them will lead to an incorrect result and decrease the final accuracy score. Also, the result of the model of Tufano et al. is very similar to the one that they reported in their papers~\cite{tufano2018empirical, tufano2019learning}, indicating that it is an expected performance for such datasets.

\begin{table}[ht]
    \caption{Evaluation Results of Applying Changes to Code}
	\begin{center}
	\begin{tabular}{ l  c }
		\toprule
		\multicolumn{1}{c}{\textbf{Model}} &  \textbf{Accuracy} \\ \midrule
		Tufano et al. & 5.9\% \\ 
		\yinmodel & 4.0\% \\ 
		\toolname & 9.9\% \\ 
		\bottomrule
	\end{tabular}
	\end{center}
	\label{tab:evaluation_of_model_modifications}
\end{table}

The model of Tufano et al. reaches an accuracy of 5.9\%. \yinmodel reaches 4.0\% accuracy, which is even less than the first baseline. From this, we conclude that edit representations from different examples might disturb the decoder from generating correct sequences. A possible reason for this is that the decoder pays a lot of attention to the unchanged context, and this context distracts the decoder.

At the same time, \toolname that stores only changed tokens in the edit sequences is able to generalize code changes better. Our model performs more accurately than \yinmodel and the model of Tufano et al., and reaches the best accuracy of 9.9\%, which is 5.9 percentage points (p.p.) greater than \yinmodel and 4.0 p.p. greater than the baseline by Tufano et al. 
From this, we conclude that edit representations built by our model guide it during the decoding step and allow to transform code fragments more accurately according to the desired change. 

To summarize, the evaluation on the \textit{Labeled} dataset shows that the proposed modifications allowed \toolname to make edit representations more generalized compared to \yinmodel.

\subsection{Commit Message Generation}

\Cref{tab:10_fold_cv_commit} shows the results for the commit message generation task: the mean values and standard deviations of BLEU scores over 10 evaluations. The results on the original dataset are from the paper of Liu et al.~\cite{liu2019generating} for both baseline models. The authors conducted their experiments on a single fold, therefore, standard deviation values are not available.

\begin{table}[h]
    \caption{Evaluation Results of Commit Message Generation (BLEU scores, percent)}
    \vspace{-0.2cm}
	\begin{center}
	\begin{tabular}{ l  l  l  l }
		\toprule
		  \multicolumn{1}{c}{\textbf{Dataset}} & \multicolumn{1}{c}{\textbf{Jiang et al.}} & \multicolumn{1}{c}{\textbf{Liu et al.}} & \multicolumn{1}{c}{\textbf{This work}} \\ \midrule
		Original & $37.0$  & $39.0$ & $39.3 \pm 1.0$ \\
		Filtered & $40.0 \pm 1.2$ & $41.9 \pm 1.0$ & $40.5 \pm 1.0$ \\ 
		\bottomrule
	\end{tabular}
	\end{center}
	\label{tab:10_fold_cv_commit}
	\vspace{-0.2cm}
\end{table}

On the \textit{Original} dataset, our approach achieves the mean BLEU value of 39.3\%, which is 2.3 and 0.3 p.p. greater than the models of Jiang et al. and Liu et al., respectively. However, taking into consideration the standard deviation of 1.0\%, we cannot reliably state that our approach performs better or worse. We assume that the models of Liu et al. and Jiang et al. have comparable standard deviations, which means that the performance of all the models should be considered equal.

On the \textit{Filtered} dataset, the model of Liu et al. has the best performance in terms of mean values (41.9\%). Our model achieves the mean BLEU value of 40.5\%, which is between the models of Liu et al. and Jiang et al. Again, we cannot reliably distinguish these results, since all the models have standard deviations that are at least 1.0\%.
As described in \Cref{sec:commit-message-generation}, \toolname was pre-trained and fine-tuned on different parts of the dataset. During the fine-tuning step, the weights of the model responsible for edit and source code representations were frozen, meaning that the feature extraction process was not fine-tuned for commit message generation process. Still, we can notice that \toolname did not lose in performance even in this setup. 

From this we conclude that general-purpose vector representations, even when trained on a different dataset, still contain the same amount of information that existing supervised models are able to extract. This means that it might be justified to pre-train the model on a much larger dataset that does not contain commit messages and then fine-tune on a small dataset that does contain commit messages. We leave this for the future work.

\subsection{Discussion}

We conducted an evaluation on two different tasks: applying edits to code fragments and summarizing code changes in natural language. Our experiments show that the proposed approach successfully learns embeddings of code changes. We conclude that \toolname not only vectorizes code changes but also preserves their semantics in the obtained representations. The model encodes edits in a generalized way, meaning that it extracts features that describe the whole change.

In deep learning, the amount of available data plays a vital role, and usually unlabeled data is easier available than labeled data. Considering the fact that \toolname learns embeddings in an unsupervised way, the proposed approach has great potential. Large pre-trained Transformer models~\cite{devlin2018bert, vaswani2017attention, kanade2019pre} have already made a breakthrough in building embeddings of text and code. In this work, we come to a conclusion that similar ideas actually can be applied to code changes, and the proposed approach can perform on par with the best supervised approaches.
\vspace{-0.2cm}
\section{Threats to Validity}\label{sec:threats-to-validity}

We only used GitHub as a source of code changes, while there are also other hosting platforms both for open-source and private projects. To minimize this threat, we evaluated the approach on several datasets that include projects with diverse topics.

Also, all the projects in our datasets are written in Java. However, our model does not employ any code features specific to Java.
Besides, we tested our approach only on two practical tasks. However, the selected tasks are complex enough to test the desired properties of the proposed approach. 
We leave further testing of our approach on other languages, datasets, and tasks for future work.

Open-source projects might contain duplicated code fragments, so we checked all the data that we used for having exact duplicates. Several researchers claimed that there are many automatically generated commit messages in the dataset by Jiang et al. that we used~\cite{liu2018neural, van2019generating}. After filtering, the models usually show a simultaneous decrease in their performance. This makes their direct comparison still possible, so we use the original dataset.

\vspace{-0.2cm}
\section{Conclusion}\label{sec:conclusion}

In this work, we propose a novel approach for building embeddings of code changes. The proposed model can be pre-trained on a large unlabeled corpus of code changes in an unsupervised manner. After pre-training, the model is able to build general-purpose distributed representations of code changes. To solve tasks that involve code changes, the model only needs to be fine-tuned on a small corpus of labeled data.

We implement the proposed approach and publish all the code and data: \url{https://zenodo.org/record/5082684}. We also replicate and publish the code of several other approaches that we use as baselines in our study~\cite{yin2018learning, jiang2017automatically, liu2019generating}. Some of their implementations were not publicly available before.

We show that embeddings of code changes built by our model are more generalized and semantically meaningful compared to the approach by Yin et al.~\cite{yin2018learning}. This helps us to solve the task of applying changes to code better, improving the accuracy by $5.9$ p.p. To demonstrate that the model can be fine-tuned for a specific downstream task, we conduct additional experiments on the commit message generation task. The generated embeddings contain enough information about a change to generate a short description of it in natural language. The performance of our model is comparable to the performance of the models, designed and trained specifically for this particular task.

\bibliographystyle{ACM-Reference-Format}
\bibliography{cites}
\end{document}